\documentclass{skads2009}
\usepackage{graphicx}

\usepackage{natbib}                     
\bibpunct{(}{)}{;}{a}{}{,}              
\begin{document}
   \title{A Digital Broadband Beamforming Architecture for 2-PAD}

   \author{R. Armstrong\inst{1}, J. Hickish\inst{1}, K. Zarb Adami\inst{1} and M.E. Jones\inst{1} }

   \institute{University of Oxford, Denys Wilkinson Building, Keble Road, Oxford, OX1 3RH, United Kingdom
   }

   \abstract{We describe an hierarchical, frequency-domain beamforming architecture for synthesising a sky beam from the wideband antenna feeds of digital aperture arrays. The development of densely-packed, all-digital aperture arrays is an important area of research required for the Square Kilometre Array (SKA) radio telescope.  The design of real-time signal processing systems for digital aperture arrays is currently a central challenge in pathfinder projects worldwide. In particular this work describes a specific implementation of the beamforming architecture to the 2-Polarisation All-Digital (2-PAD) aperture array demonstrator.\\
   
   }
   \authorrunning{Armstrong et. al.}
   \maketitle
%
%
\section{Introduction}
The development of densely-packed, all-digital aperture arrays is an important area of research required for the Square Kilometre Array (SKA) radio telescope. The design of real-time signal processing systems for aperture arrays is certainly one of the most challenging tasks in the design of next-generation instruments. In this paper, we describe the design and implementation of an hierarchical beamforming architecture for wideband radio telescopes.\\

\begin{figure}
\centering
\includegraphics[width=85mm]{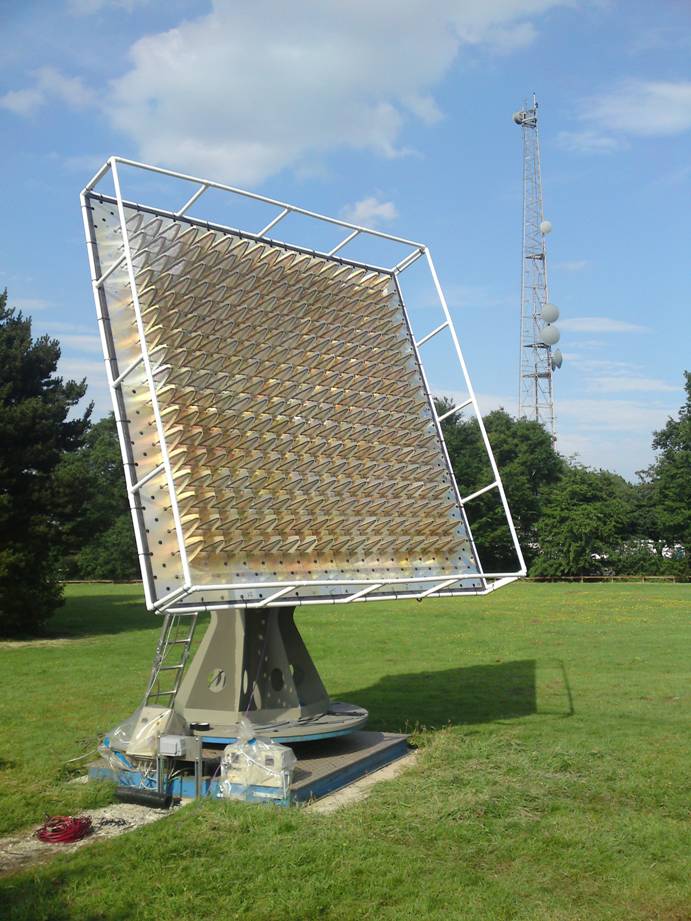}
\caption
	{The 2-Polarisation All-Digital (2-PAD) aperture array demonstrator at Jodrell Bank.} 
\label{2PAD}
\end{figure}

This document is organised as follows. Chapter I provides an introduction to the problem in question and Chapter II describes the theory and process of `beamforming' for aperture arrays used in radio astronomy. Chapter III describes the design and implementation of a frequency-domain beamformer on the dual-polarisation all-digital aperture array demonstrator at Jodrell Bank (see figure \ref{2PAD}). We begin this chapter by describing the incremental design of a 4-element system and then describe the move to 16 (4x4) elements. Chapter IV  describes calibration procedures for the array, and is followed by discussion and conclusions.\\

\subsection{Aperture Phased Arrays}
\begin{figure}
\centering
\includegraphics[width=70mm]{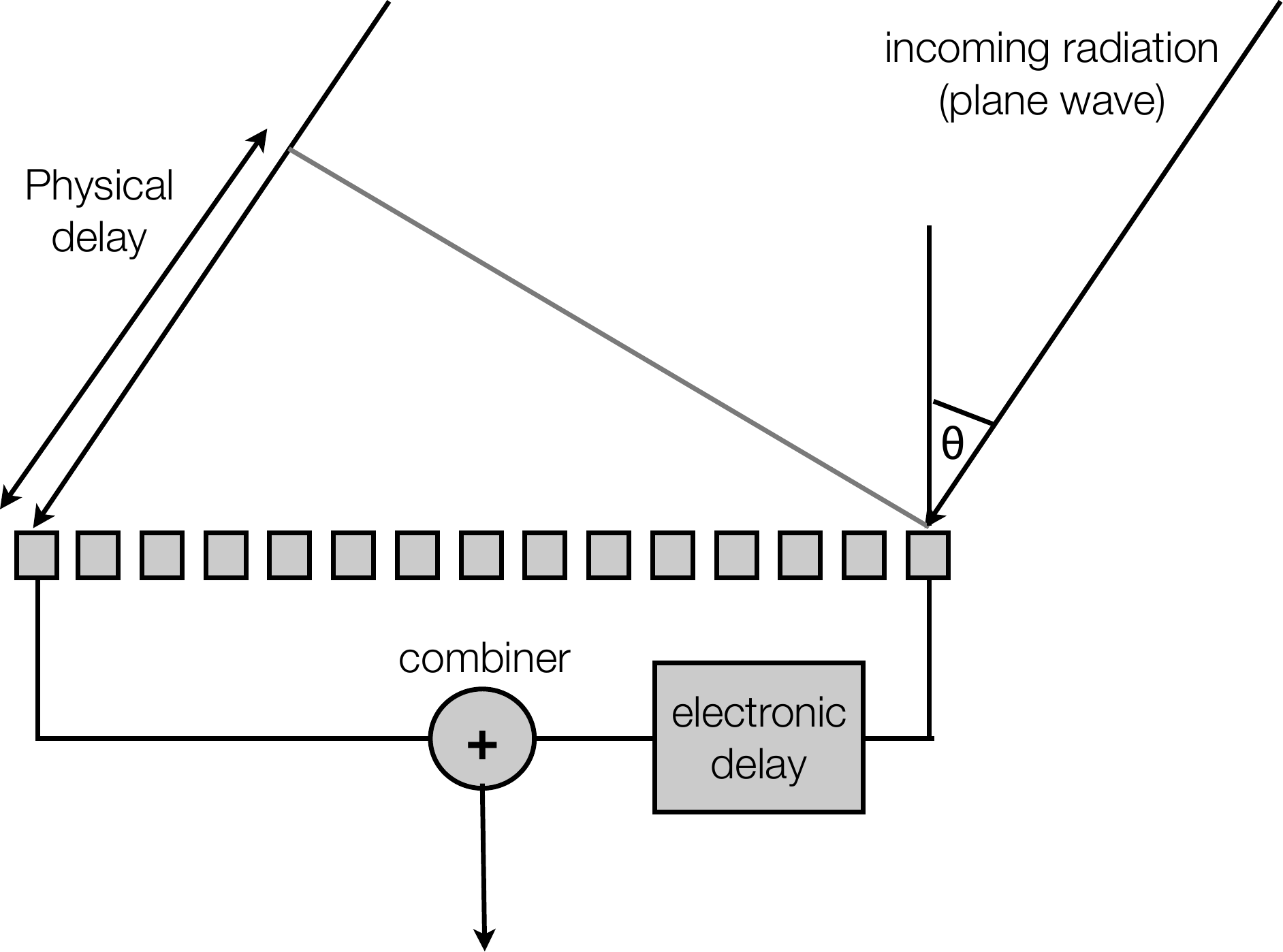}
\caption
	{An illustration of the electronic processing of signals received from an array of antennas for concentrating the signal from a certain direction. This is the basic principle of the phased aperture array: different delay configurations allow beams to be formed for different values of \(\theta\)} 
\label{BeamformingIllustration}
\end{figure}

An aperture phased array is a collection of fixed, non-directional antennas whose outputs are combined to synthesise an effective directional antenna. This synthesised antenna is pointed electronically; significantly the physical component antennas do not ever move. The electronic combination of signals with beamforming techniques performs the same spatial filtering task as a parabolic dish would in a traditional radio telescope dish. Figure \ref{BeamformingIllustration} illustrates the basic principle of this operation.\\

The flexibility, survey speed, system-level autonomy and scalability provided by antenna arrays make them extremely attractive for next-generation radio instruments. This publication discusses their use in radio astronomy, but the techniques are applicable to general wideband detection and communication instruments. Future radio instruments will certainly be dependent on the availability of high-performance signal processing systems such as these. Since aperture arrays are seen as a critical antenna technology to achieve the required sky survey speed in future radio telescopes (see, for example, \cite{2006SKA.memo.81}), the development of a scalable signal processing system is of great importance.


\section{Beamforming}

\subsection{Digital, Frequency-Domain Beamforming}
The process of beam forming, shaping and summing of tile elements may be performed in either the digital or analogue domains. Digital beamformers are distinguished from analogue and partially analogue beamformers by signal digitisation directly after initial RF amplification, followed by processing entirely in the digital domain. The all-digital approach offers maximum flexibility in manipulation of incoming signals both in beamforming and in calibration of the wideband signals. However, there are many tradeoffs in this space, and different applications will dictate eventual implementation.\\

Beamforming techniques can be categorised as time-domain or frequency domain. For broadband applications, sub-banded frequency-domain techniques may be seen as superior due to their computational advantage if fine sample interpolation operation is implemented (see, for example, \cite{1027910}) and the ability to calibrate band-pass characteristics of the analogue system (see \cite{priv:2009}) by applying a phase shift to each sub-band in the transform domain. For these reasons, we choose to perform the beamfoming in Fourier space.



\subsection{Signal Processing and Dataflow Requirements for Wideband Phased Aperture Arrays}
In this section we describe the signal processing and dataflow requirements of the narrowband phase shift beamforming architecture. \\

The sampled broadband signal from each antenna is broken up into many narrow frequency bands. Each band is independently processed (delayed and combined with the signals from other antennas) with a narrowband phase-shift beamformer. The results of these narrowband beamformers are combined to form a broadband beam in the required spatial direction. This operation is repeated for each beam that is formed. Channelisation or `frequency binning' of large-bandwidth input signals for beamforming can be performed with various techniques.\\

\begin{figure}
\includegraphics[width=80mm]{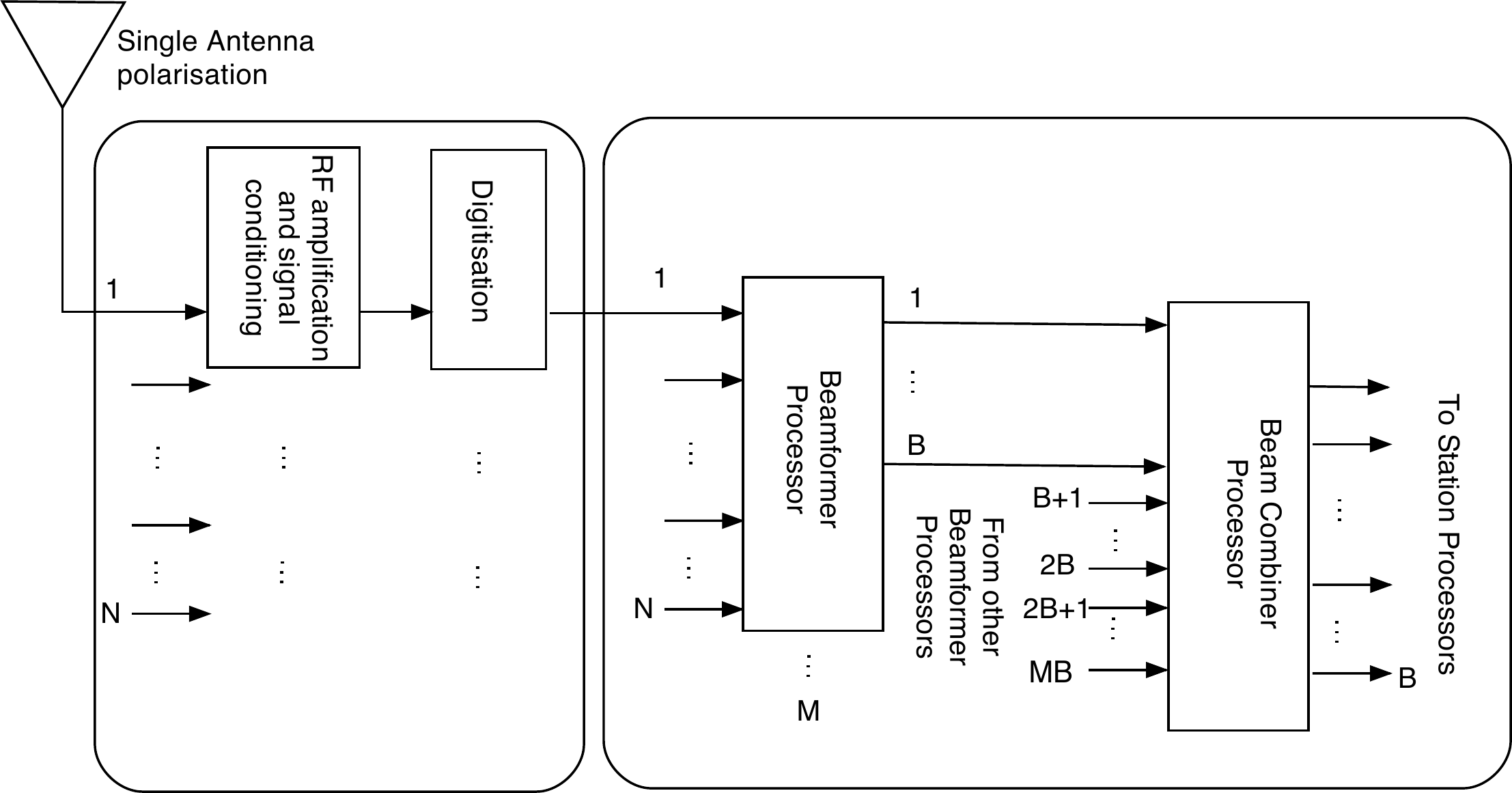}
\caption[System Signal Processing Flow for Digital Aperture Array Beamforming]
	{System Signal Processing Flow for Digital Aperture Array Beamforming at Tile Level} 
\label{SPFlowAA}
\end{figure}

Figure \ref{SPFlowAA} shows the signal processing flow for the phase-shift beamforming architecture. N single-polarisation antenna signals are digitised then combined with phase correction factors to form B beams in the beamformer processor. These beams are combined in a beam-combiner processor\footnote{The beam-combiner processor may be a further hierarchy of processors}. The resultant `station beams' (which are wideband spectra) are then sent to a central correlator for interferometric processing.\\

We now analyse the dataflow requirements: the analogue input signal ranges from 0.5GHz to 0.7GHz (a bandwidth of 0.2GHz), and is sampled with $q$ bits of precision, which results in a digital input bandwidth from each antenna polarisation of $q$ *0.4Gbps. This means that the first-stage beamformer processor will have a raw input data-rate of $Nq*0.4$ Gbps and an output data rate of $Bqs$*0.4 Gbps, where $s$ is the bitwidth scaling factor of the beamformer.\\

Note that there is a strong distinction between high-performance, real-time processing which must be performed at the input data-rate and more computationally complex co-efficient calculation which may be performed at a much reduced rate. In this paper, we concentrate on the former; this is an implementation of the high-performance processing system. We allow the steering and correction coefficients to be updated at a lower rate, dictated by the speed with which the array must be able to scan the sky.

\section{An Hierarchical Frequency Domain Beamforming Architecture}
In this section, we describe the design and implementation of an FPGA-based digital system for the frequency sub-band phase-shift beamforming architecture.\\

\begin{figure}
\begin{center}
\includegraphics[width=70mm]{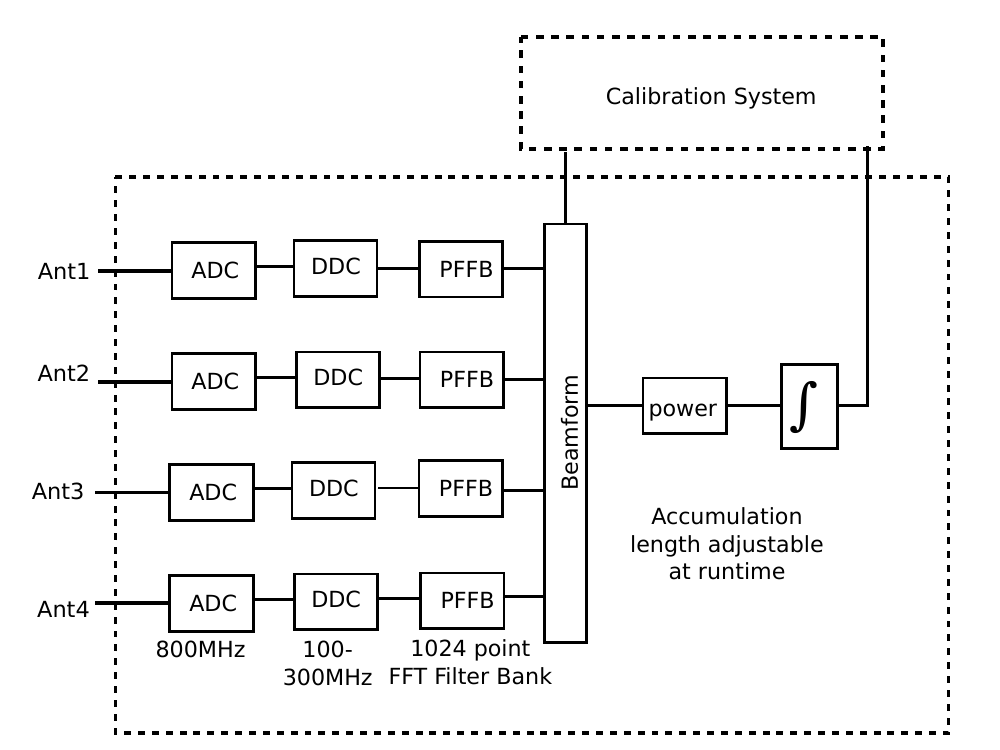}
\caption{An architectural drawing of the 4-element beamformer and calibration system} 
\label{2pad_arch}
\end{center}
\end{figure}

Figure \ref{2pad_arch} is an architectural schematic of the 4-element beamformer digital system. The Analogue to Digital Converter (ADC) samples the incoming antenna signal at 800MSa/s with 8 bits of precision. This 400MHz bandwidth signal is immediately Digitally Down-Converted (DDC) by mixing with an complex sinusoid at 3/4 of the ADC clock rate to yield a base-banded 500MHz to 700MHz input signal. Channelisation is the next step in the processing pipeline: an efficient technique is the Polyphase FFT\footnote{Fast Fourier Transform (FFT)} Filter Bank (PFFB) channeliser, which improves frequency isolation of the FFT operation (see for example \cite{2008PASP..120.1207P}) and has shown to be efficient for large numbers of equally spaced channels \citep{Pucker01}. The beamformer applies steering and correction coefficients to each signal stream and then performs beam summation. We then accumulate the power sum of the raw voltage beam signal for a run-time configurable length of time.\\

We have implemented this design with signal processing libraries and hardware designed by the Collaboration for Astronomy Signal Processing Research (CASPER)\footnote{for more information, see \textsf{www.casper.berkeley.edu}}. We are indebted to all collaborators of this group for the support they have provided with hardware and libraries.\\

The FPGA-based design uses embedded multipliers in a Xilinx Virtex II Pro device. Since these fixed hardware blocks are natively 18-bit, we preferentially use 18-bit multiplications and datapaths throughout the design.

\subsection{Extension to 16 Antennas}
\begin{figure}
\centering
\includegraphics[width=70mm]{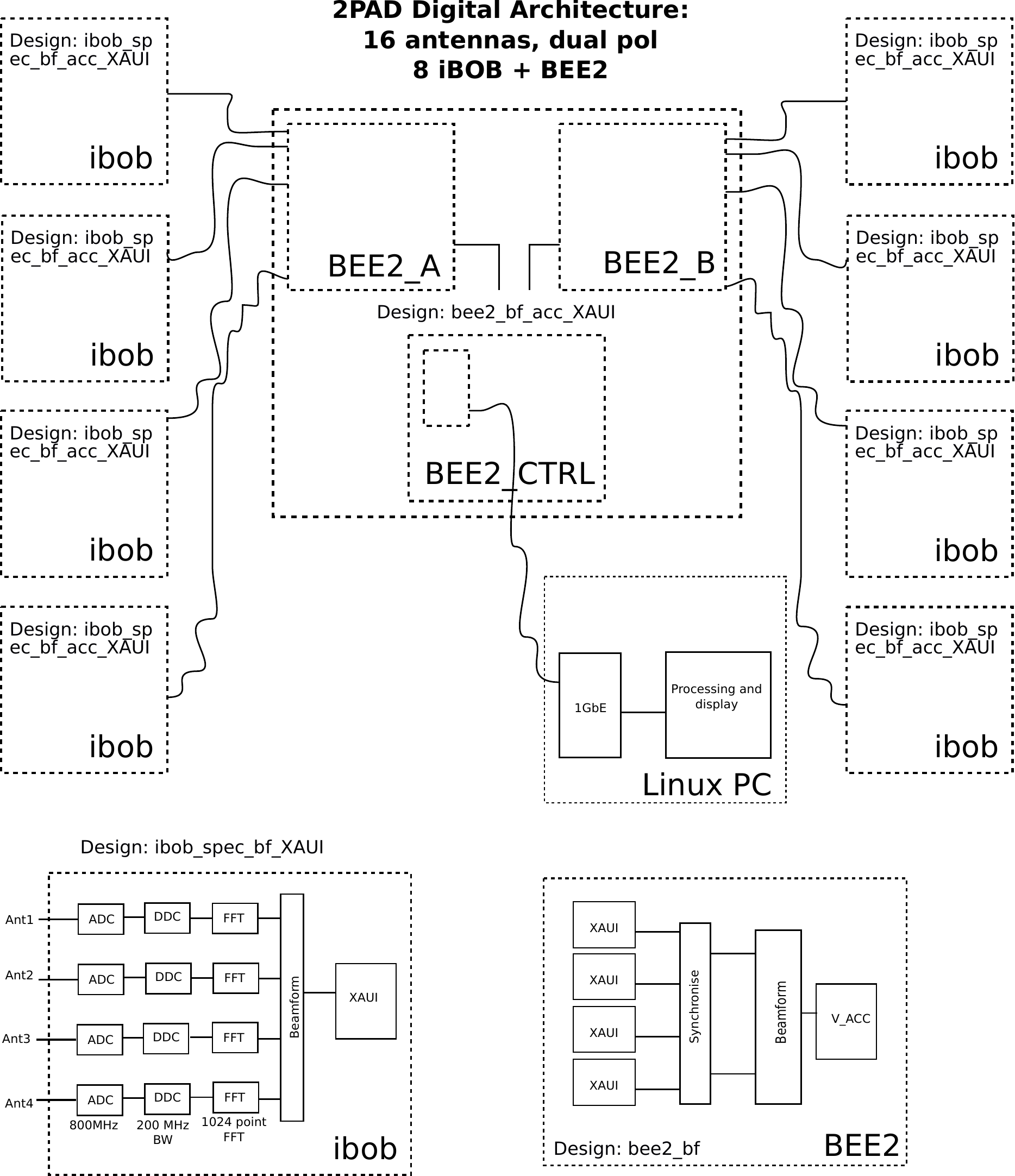}
\caption
	{A schematic representation of the 4x4x2 digital beamformer showing communication and synchronisation between the iBOB boards and the BEE-2 boards.} 
\label{4x4x2_arch}
\end{figure}

In order to scale the beamformer to more receiving elements communication between digital boards is necessary. We have chosen to use the XAUI protocol to accomplish this task. A schematic representation of the 4x4-element dual polarisation beamformer design is shown in figure \ref{4x4x2_arch}. The system is a synchronous design, implemented with 8 iBOB boards and using 2 of the 4 `corner' FPGAs of the BEE2 board.

\section{Calibration at Aperture Array Tile Level}

The full digital calibration of close-packed, strongly electromagnetically coupled antenna arrays has been shown to be a challenging problem. There is great need for the development of a robust calibration scheme in order to achieve the high dynamic range required of digital aperture arrays.\\

Calibration of individual elements of a dense aperture array at the tile level is fundamentally different from traditional radio telescope calibration for two reasons. Firstly, individual antennas `see' the entire observable sky; the dish-centric concept of Field of View (FoV) at the antenna level is not valid. Secondly, each individual element does not have the sensitivity to detect astronomical sources.\\

Thus we must approach the calibration in other ways. Perhaps the obvious first suggestion would be to fully constrain the analogue system. Certainly, this is done to the first degree. We refer the reader to \cite{priv:2009} for discussions about analogue signal chains, including S-parameter propagation, variation due to temperature, component anisotropy and radio-frequency interference (RFI) effects. That is, a typical analogue system is frequency, temperature, signal chain, component, RFI-environment and pointing angle dependent, and needs to be continuously calibrated on timescales faster than the fastest varying of the set of all these.\\

A first approach to this calibration problem would be the correlation approach. At a minimum, a correlation-based technique must perform an Nx1 correlation. A full correlation of all baselines would allow a over-constrained set of linear equations to be formed, providing a robust solution for gain and phase of each element. In this case, full broadband signal path calibration is possible with any signal, including noise. However, a real-time correlator may require an even larger digital system than the entire beamformer. Thus, correlation approaches may be required to be time multiplexed.\\

\begin{figure}
\centering
\includegraphics[width=90mm]{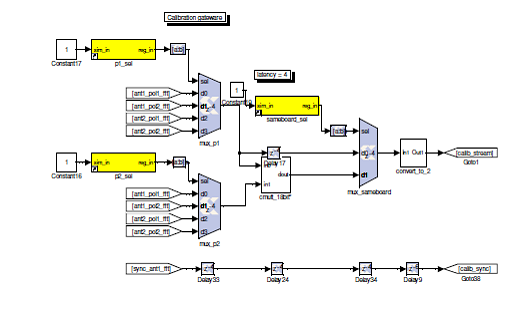}
\caption
	{A design representation of the correlation-based calibration block. It determines the RF path lengths of signals before the beamforming procedure. As shown in the figure, the calibration forms the complex correlation between a reference signal and the target signal.} 
\label{Calibration}
\end{figure}

Another possible solution is to use the existing beamformer architecture to calibrate off a strong signal. This signal may be externally injected into the analogue system either directly with a noise diode power-combiner scheme or with a known strong signal beamed at the entire array.\\

We initially chose the latter for our calibration, beaming a 675MHz Continuous-Wave (CW) source at the antenna array from the far field. We subsequently implemented a time-multiplexed correlation-based array calibration, the design of  which is shown in figure \ref{Calibration}




\section{Results and Discussion}

\subsection{Anechoic Chamber Testing}
\begin{figure}
\begin{center}
\includegraphics[width=90mm]{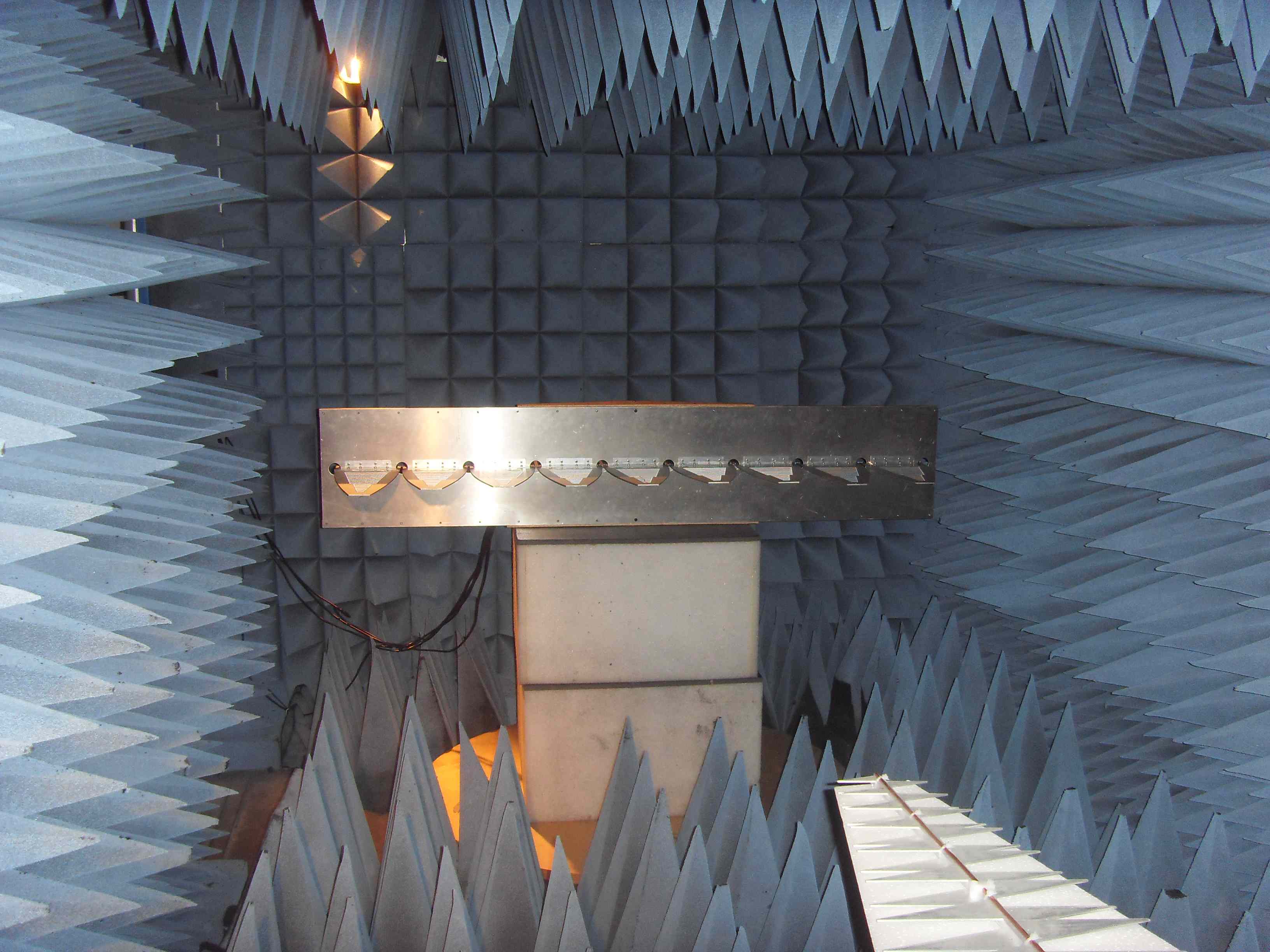}
\caption
	{An image showing the four-element antenna array in the anechoic chamber.} 
\label{AC}
\end{center}
\end{figure}

We have tested the 4-element beamformer using the power-optimisation calibration procedure in a radio-frequency anechoic chamber at the Department of Engineering at Oxford University. Significantly, the beamformer is tested \emph{without} the analogue chain that is required in the eventual radio instrument in the field. Figure \ref{AC} shows the linear, four-element antenna array in the chamber. The transmit antenna is located in the far-field of the receive array so we neglect the transmit aperture illumination function.


\subsection{Field Deployment}
This calibration process has also been trialled in the field. A transmitter is positioned atop a mast (again, in the far-field of the receive array) and pointed at the array. However, in this case, we included the analogue chain to the antennas before digitisation. The calibration procedure was performed in the same way as before. Figure \ref{ACvsField} shows a 700MHz beam at 0 degrees obtained in the anechoic chamber overlaid with a beam at the same frequency obtained in the field. \\

\begin{figure}
\begin{center}
\includegraphics[width=90mm]{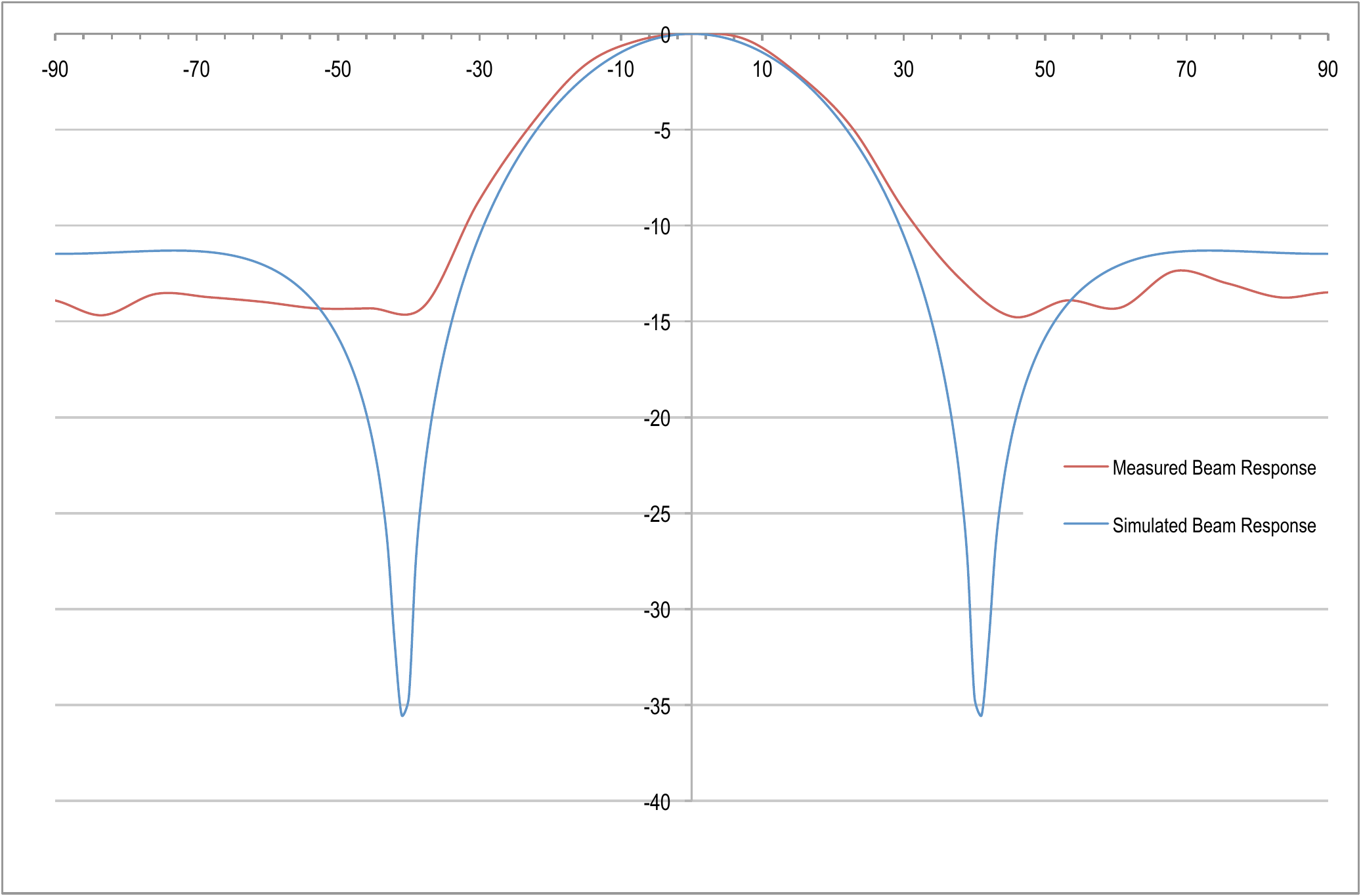}
\caption
	{A plot of the sixteen-element (4x4) beam pattern of the array at 675MHz as measured in the field at Jodrell Bank plotted alongside a simulated (ideal) 4x4-element beam. In this figure the beam was calibrated while pointed toward the transmitter.  The abscissa is the scan angle in degrees from broadside and the ordinate is an arbitrary power scale relative to maximum signal power at the calibrated 0 degree pointing.} 
\label{4x4x1_1}
\end{center}
\end{figure}

\begin{figure}
\begin{center}
\includegraphics[width=90mm]{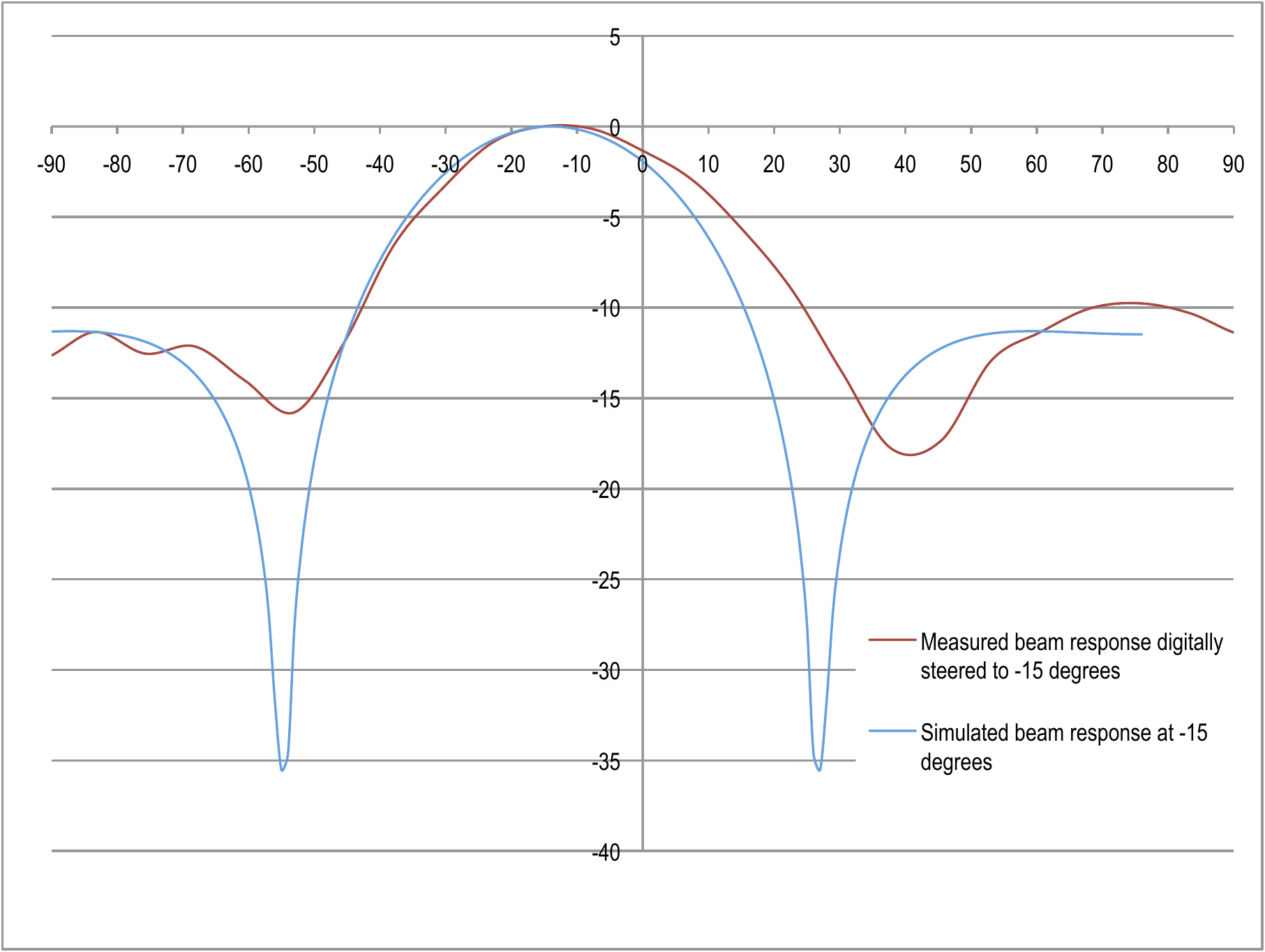}
\caption
	{A plot of the sixteen-element (4x4) beam pattern of the array at 675MHz as measured in the field at Jodrell Bank with the array physically steered approximately 15 degrees away from the signal source. Plotted alongside for comparison is a simulated (ideal) 4x4-element beam pointed directly at the source. In this figure the array was first calibrated while pointed toward the transmitter, then mechanically steered approximately 15 degrees off the source. The abscissa is the scan angle in degrees from broadside and the ordinate is an arbitrary power scale relative to maximum signal power at the calibrated 0 degree pointing.} 
\label{4x4x1_2}
\end{center}
\end{figure}

Further to this, a 4x4 array was installed in the field at Jodrell Bank Observatory and beam measurements taken. These beams are shown in figures \ref{4x4x1_1} and \ref{4x4x1_2}. We expect that the `field' beam will be corrupted by effects like structure scattering\footnote{where structure scattering includes signal reflection off array structures, ground planes and surrounding vegetation} and the various analogue system effects. Figure \ref{ACvsField} clearly shows that field calibration and measurement results in a beam with significant error compared to the beam obtained in the anechoic chamber. In figure \ref{4x4x1_2}, the response indicates the relative direction to the source when the array is steered away mechanically. Even though the resolution of the measurement taken with 16 elements is less than that of the 4-element measurement, it can be seen that noise immunity is far better with the larger array. 

\section{Conclusions}

In this paper we propose that digital aperture arrays are a fast, useful and elegant collector technology for GHz radio astronomy, that the development of high-performance signal processing systems is critical to the success of next-generation radio instruments, and have described the design of a dual-polarisation, 16-element frequency-domain aperture array beamformer, which was deployed on the 2PAD instrument at the Jodrell Bank Observatory.


\bibliographystyle{plainnat}
\bibliography{bibChapter1}







%
%





\end{document}